\documentclass{article}
\usepackage{graphicx}
\begin{document}

\title{Mesoscopic multiterminal Josephson structures. \textsl{I}.
Effects of nonlocal weak coupling.}

\author{M.H.S. Amin$^{a}$, A.N. Omelyanchouk$^{b}$, and A.M.
Zagoskin$^{a,c}$ \\
{\small {\em $^{a}$D-Wave Systems Inc., 320-1985 W. Broadway,}}\\
{\small {\em Vancouver, B.C., V6J 4Y3, Canada}} \\
{\small {\em $^{b}$B.I.Verkin Institute for Low Temperature
Physics and Engineering,}} \\
{\small {\em Ukrainian National Academy of Sciences,
Lenin Ave. 47, Kharkov 310164, Ukraine }}\\
{\small {\em $^{c}$Physics and Astronomy Dept., The University of
British Columbia,}}\\
{\small {\em 6224 Agricultural Rd., Vancouver, B.C., V6T 1Z1,
Canada}} \date{}}\maketitle

\begin{abstract}
We investigate nonlocal coherent transport in ballistic
four-terminal Josephson structures (where bulk superconductors
(terminals) are connected through a clean normal layer, e.g., a
two-dimensional electron gas).

Coherent anisotropic superposition of macroscopic wave functions
of the superconductors in the normal region produces phase slip
lines (2D analogs to phase slip centres) and time-reversal
symmetry breaking 2D vortex states in it, as well as such effects
as phase dragging and magnetic flux transfer. The tunneling
density of local Andreev states in the normal layer was shown to
contain peaks at the positions controlled by the phase
differences between the terminals.

We have obtained general dependence of these effects on the
controlling supercurrent/phase differences between the terminals
of the ballistic mesoscopic four-terminal SQUID.
\end{abstract}
\newpage

\section{Introduction.}
The multiterminal Josephson  junctions \cite{lik,kok} generalizes
the usual (two-terminal) Josephson junctions \cite{bar} to the
case of weak coupling between several massive superconducting
banks (terminals). Compared with two-terminal junctions, such
systems have additional degrees of freedom and the corresponding
set of control parameters, preset transport currents and (or)
applied magnetic fluxes. As a result, the current- or
voltage-biased and the magnetic flux-driven regimes can be
combined in one multiterminal microstructure.

One of the realizations of multiterminal coupling is a system of
short dirty microbridges going from a common center to separate
massive superconductors. The theory of this kind of
multi-terminals was derived in \cite{VO1,oov} within the
phenomenological Ginzburg-Landau scheme (Aslamazov and Larkin
model \cite{AL}). This approach is valid for temperatures $T$
near the critical temperature $T_{c}$ and for the local case when
the characteristic spatial scale is larger then the coherence
length $\xi_{0}\sim \hbar v_{F} /T_{c}$. The stationary states
and the dynamical behaviour of the microbridge type
multiterminals were studied for different microstructures,
four-terminal SQUID controlled by the transport current, weakly
coupled superconducting rings (see review of theoretical and
experimental results in \cite{oo,vlem}).

The Josephson effect in mesoscopic weak links with direct
conductivity (S-N-S junctions  , ballistic point contacts)
exhibits specific features \cite{SNS},\cite{KO} which are absent
in conventional dirty microconstrictions near $T_{c}$ \cite{AL}.
As in normal metal mesoscopic structures \cite{Imry}, the
electrodynamics of supercurrents in mesoscopic Josephson
junctions is nonlocal. Supercurrent density depends on the
spatial distribution of the superconducting order parameter in
all points of the mesoscopic weak link region. The coherent
current flow is carried by the Andreev states \cite{Andreev}
formed inside the weak link. Nonlocal nature of mesoscopic
supercurrents was demonstrated by Heida {\it et al.}\cite{Klap},
investigating the mesoscopic S-2DEG-S (superconductor-two
dimensional electron gas - superconductor)  Josephson junctions.
They measured $2\Phi_0$ periodicity of the critical current
instead of the standard $\Phi_0$ ($\Phi_0=hc/2e$ is the magnetic
flux quantum). Theory of this effect was developed in
Refs.\cite{Zag,Leder}.

The current level of nanofabrication technology made it possible
to realize multi-terminal mesoscopic Josephson junction similar
to the 2-terminal junction studied in \cite{Klap}. The microscopic
theory of the mesoscopic ballistic Josephson multi-terminals was
derived in Ref.\cite{malekom}. It is valid for arbitrary
temperatures $0<T<T_c$ and describes the nonlocal coherent
current states in the system. The effects of nonlocal coupling,
such as phase dragging and magnetic flux transfer were obtained
in Ref.\cite{ommalek}.

    In the present paper we continue study of quantum
interference effects in mesoscopic multiterminals, which are
related to the nonlocality of weak coupling. The paper consists
of two parts. In first part (Article I) the effects of nonlocal
coupling in mesoscopic multiterminal structures are studied. The
general properties of Josephson multiterminals are described in
Section 2. Section 3 gives the results concerning the  current
distribution and density of states inside weak link. In Section 3
we study specific for mesoscopic case properties of four-terminal
SQUID. In the second part (Article II) a superconducting phase
qubit based on mesoscopic multiterminal junction is proposed and
investigated.

\section{Mesoscopic Four-Terminal Junction.}

\subsection{System description}

In a mesoscopic 4-terminal junction, the bulk superconductors
(terminals) are weakly coupled to each other through a clean
two-dimensional normal metal layer (2D electron gas) as it is
shown in Fig.1.
\begin{figure}[!t]
\begin{center}
\includegraphics[width=3in , height=3in]{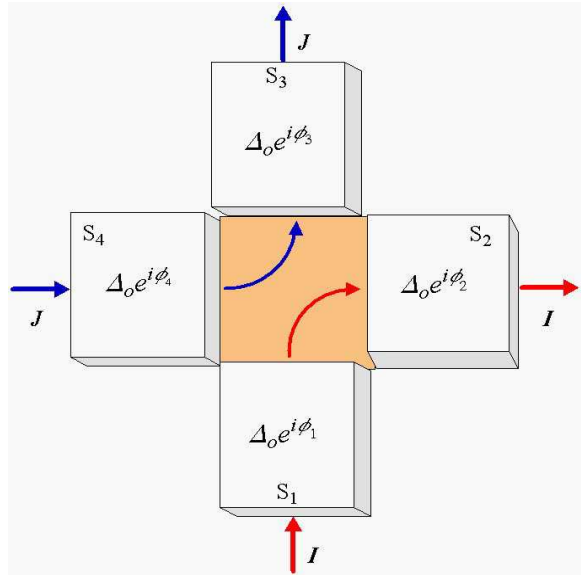}
\end{center}
\caption{{Mesoscopic four-terminal Josephson junction with
``parallel" implementation of the supercurrents. The four bulk
superconducting regions, S$_1$...S$_4$, are weakly coupled through
the thin layer of normal metal (2DEG), represented by the shaded
area.}} \label{4tm}
\end{figure}
The pairs of terminals can be incorporated in bulk
superconducting rings or in circuits with preset transport
currents. In Fig.2 we show two such configurations. The first one
(Fig.2a) presents two superconducting rings, each interrupted by
a Josephson junction, which are at the same time weakly coupled
to each other. The second configuration (Fig.2b), combines a
current (or voltage) biased junction and a flux driven junction
in the ring. We call this configuration the 4-terminal SQUID
controlled by the transport current.

\begin{figure}[!t]
\begin{center}
\includegraphics[width=0.8 \textwidth]{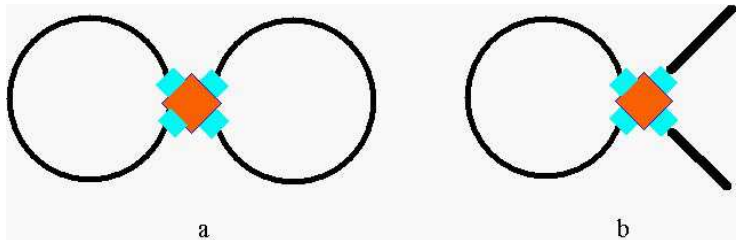}
\end{center}
\caption{{Superconducting microstructures based on mesoscopic
four-terminal Josephson junctions. (a) Two weakly coupled
superconducting rings. (b) Mesoscopic four-terminal SQUID. }}
\label{squidring}
\end{figure}

The state of the \textit{i}-th terminal S$_i$ (\textit{i}=1...4)
is determined by the phase $\varphi_{i}$ of the complex
off-diagonal potential $\Delta_{0}\exp({\rm i}\varphi_i)$.
Superconducting banks induce the order parameter $\Psi$ in the
normal metal region (shaded area in Fig.1). Inside this
mesoscopic, fully phase coherent weak link, the supercurrent
density ${\bf j} (\vec \rho)$ at point $\vec \rho$
\textsl{nonlocally} depends on the values of the induced order
parameter $\Psi$ at all points ${\vec \rho}\ '$. In its turn, the
order parameter $\Psi(\vec \rho)$ depends on the phases
$\varphi_i$. The total current $I_i$, flowing into $i$-th
terminal depends on the phases $\varphi_j$ of all the banks and
has the form \cite{malekom} :
\begin{equation}
I_{i}= \frac{\pi \Delta_0 }{e} \sum_{j=1} ^{4} \gamma_{ij}
\sin{\left( \frac{\varphi_{i}-\varphi_{j}}{2} \right) }\tanh{
\left[ \frac{\Delta_0
\cos{(\frac{\varphi_{i}-\varphi_{j}}{2})}}{2T} \right] }.
\end{equation}
In the case of two terminals Eq. (1) reduces to the formula for
ballistic point contact \cite{KO} with $\gamma_{12}$ equals to
Sharvin's conductance.

Expression (1) corresponds to the case of small junction, when
linear dimensions of the N-layer are smaller than the coherence
length $\xi \sim \hbar v_F/\Delta_0$ (for the case of arbitrary
junction's dimensions see Ref.\cite{malekom}). We are focusing
here on the small junction case because the effects of
nonlocality are most pronounced in this situation. The geometry
dependent coefficients $\gamma_{ij}$ denote the coupling between
the partial Josephson currents in ballistic  two-terminal
S$_i$-S$_j$ weak links.

Eq. (1) is simplified when $T \approx 0$, or when $T \approx T_c$.
%
In $T=0$ limit, it becomes
\begin{equation}
I_i = \frac{\pi \Delta_0(0) }{e} \sum_{j=1}^{4} \gamma_{ij} \sin
\left( {\varphi_i-\varphi_j \over 2} \right) {\rm sign} \left[
\cos \left( {\varphi_i-\varphi_j \over 2} \right) \right].
\end{equation}
Near $T_c$ on the other hand, the order parameter is small,
$\Delta_0 \rightarrow 0$ and one can write
\begin{equation}
I_i =  \frac{\pi {\Delta_0(T)}^2}{4 e T_c} \sum_{j=1} ^{4}
\gamma_{ij} \sin (\varphi_i-\varphi_j).
\end{equation}
Equations (2) and (3) are qualitatively similar, differing by the
magnitude of critical currents and by the shape of current-phase
dependencies ($ \sin(\frac {\varphi} {2}){\rm sign} [ \cos(\frac
{\varphi}{ 2})] $ and $\sin(\varphi)$). For definiteness, in the
following we will consider the case of $T \sim T_c$, keeping in
mind that the results hold qualitatively at low temperatures as
well.

For the Josephson coupling energy of the junction $E_J$ , which
relates to the supercurrents $I_i$ (3)  through
$I_i=(2e/\hbar)\partial E_{J}/\partial \varphi_i$, we have
\begin{equation}
E_J({\varphi_{i}})=\frac{\hbar}{2e} \frac{\pi {\Delta_0(T)}^2}{4
e T_c} \sum_{j<k} {\gamma_{jk}}
[1-\cos{(\varphi_{j}-\varphi_{k})}].
\end{equation}

Expression (1) for supercurrents $I_i$ looks similar to Buttiker's
multiprobe formula \cite{buttiker}
\begin{equation}
  I_i=e\sum_j \it{T}_{ij}(\mu_i-\mu_j)
\end{equation}
which relates the currents to the voltage drops between terminals
in mesoscopic normal metal multiterminal system. The similarity
reflects the above mentioned nonlocality of mesoscopic transport
on the scale of $\xi_T \sim \hbar v_F/T$ (in the ballistic limit
we are considering). The essential difference between (1) and (5)
is that unlike Josephson currents of (1), the normal currents of
(5) can flow only out of equilibrium; while the current-phase
dependence in (1) is periodic, the current-bias dependence of (5)
is certainly not.

\subsection{Circuit implementations of 4-terminal junction.
Nonlocal weak coupling.} The current-phase relations (3)
determine the behaviour of the system in the presence of the
transport currents and/or the diamagnetic currents induced by the
magnetic fluxes through the closed superconducting rings. It is
necessary to distinguish two types of circuit implementation of
the mesoscopic 4-terminal junction \cite{ommalek}. The first one,
is the ``crossed" or ``transverse" implementation, when the total
current in one circuit goes in and out through one pair of
opposite banks in Fig.1 and in the second circuit - through the
other pair. In the ``parallel" implementation, shown in Fig.1, the
currents $I$ and $J$ flow through the pairs of adjacent banks. In
this case, nonlocal coupling of currents inside the mesoscopic
N-layer results in peculiar effect of ``dragging" of the phase
difference between one pair of terminals by the phase difference
between another pair of terminals \cite{ommalek}. In the
following, we consider the ``parallel" implementation and study
the manifestations of the phase dragging effect.

The coefficients $\gamma_{ij}$ in (3,4) depend on the geometry of
the weak link (the shape of the N-layer) and on the transparency
of S-N interfaces. In general we have $\gamma_{ij}=\gamma_{ji}$
and $\gamma_{ii}=0$. For the case of parallel implementation, the
elements $\gamma_{1 2}$ and $\gamma_{3 4}$ are related to the
critical currents of the individual sub-junctions S$_1$-S$_2$ and
S$_3$-S$_4$ respectively. The matrix
\begin{equation}
\hat \gamma_{\rm coupl}=\left( \matrix {\gamma_{13}&\gamma_{14}\cr
\ \gamma_{23}&\gamma_{24}\cr}\right)
\end{equation}
describes the coupling between these two junctions. Properties of
the system (in particular the existence of the phase dragging)
qualitatively depends on whether $det(\hat \gamma_{\rm coupl})$
equals to zero or not (see Appendix). In case of a conventional
non-mesoscopic 4-terminal Josephson junction the coefficients
$\gamma_{ij}$ factorize, $\gamma_{ij}\sim (1/R_i)(1/R_j)$, where
$R_i$ are the normal resistances of dirty microbridges \cite{oo}.
This yields $det(\hat \gamma_{\rm coupl})\equiv 0 $, which we
call local coupling. On the other hand, in a mesoscopic system,
even in a completely symmetric case of $a \times a$ square
N-layer and ideal transparency ($D=1$) of N-S$_i$ interfaces, the
coefficients $\gamma_{ij}$ are given by \cite{ommalek}:
\begin{equation}
\gamma_{12}= \gamma_{34}=\gamma_0 , \quad \hat \gamma_{\rm
coupl}=\gamma_0 \left( \matrix {\sqrt{2}&1 \cr \ 1&\sqrt{2}
\cr}\right), \quad \gamma_{0} = \frac {e^2 p_{F} a}
{\sqrt{2}\hbar^2\pi^2} \left(1- {1 \over \sqrt{2}}\right),
\label{gamma}
\end{equation}
with $ det(\hat \gamma_{\rm coupl})\neq 0 $. In more general case
than the completely symmetric one (Eq. (7)), we can write
$\gamma_{ij}$ in the form
\begin{equation}
\gamma_{34}=\kappa \gamma_{12}\ , \quad \hat \gamma_{\rm
coupl}=\left( \matrix {p & q \cr q & p \cr}\right).
\end{equation}
This corresponds to a square N-layer, with different
transparencies for junctions S$_{1}$-S$_{2}$ and S$_{3}$-S$_{4}$
and/or different width of the superconductor banks connected to
the normal layer. In our numerical calculations we will use the
simple form (7), i.e. $\kappa=1$, $p=\sqrt{2}$, $q=1$.

\subsection{Current-phase relations. The phase dragging effect.}

Let us introduce new variables:
\begin{eqnarray}
\varphi _{2}-\varphi _{1} =\theta &,& \quad
\varphi_{3}-\varphi_{4}=\phi \nonumber \\
\frac{1}{2}(\varphi _{1}+\varphi _{2}) =\alpha &,&
\quad \frac{1}{2}(\varphi _{4}+\varphi _{3})=\beta \nonumber \\
\alpha -\beta =\chi &,& \quad \alpha +\beta =\gamma .
\end{eqnarray}
Without loss of generality, we can choose the phase $\gamma$
equal to zero ($\sum_{j}\phi _{j}=0$). For the circuit
implementation shown in Fig.1, we have
\begin{equation}
I=I_2=-I_1, \ \ \ J=I_3=-I_4.
\end{equation}
In terms of phase differences (9) the currents $I$ and $J$ have
the form
\begin{equation}
I=\sin{\theta}+
\left[(p+q)\sin{\frac{\theta}{2}}\cos{\frac{\phi}{2}}+
(p-q)\cos{\frac{\theta}{2}}\sin{\frac{\phi}{2}}
\right]\cos{\chi}, \label{I}
\end{equation}

\begin{equation}
J=\kappa \sin{\phi}+
\left[(p+q)\sin{\frac{\phi}{2}}\cos{\frac{\theta}{2}}+
(p-q)\cos{\frac{\phi}{2}}\sin{\frac{\theta}{2}}
\right]\cos{\chi}. \label{J}
\end{equation}
All $\gamma$'s (8) are normalized by $\gamma_{12}$ and the
currents $I$,$J$ are measured in units of
$I_0=\pi\gamma_{12}{\Delta_0(T)}^2/4eT_c$.

From the current conservation (Eq.(10)), it follows that the phase
$\chi$ in Eqs. (11) and (12) can take only two values, $0$ or
$\pi$. Minimization of $E_J$ (4) with respect to $\chi$ also gives
$\chi=0$ or $\pi$, depending on the equilibrium values of $\theta$
and $\phi$ (see Appendix):
\begin{equation}
\cos{\chi}={\rm sign} \left[(p+q)\cos{\frac{\phi}{2}}
\cos{\frac{\theta}{2}}-(p-q)\sin{\frac{\phi}{2}}
\sin{\frac{\theta}{2}} \right].
\end{equation}
\begin{figure}[t]
 \begin{center}
 \includegraphics[width=0.8\textwidth]{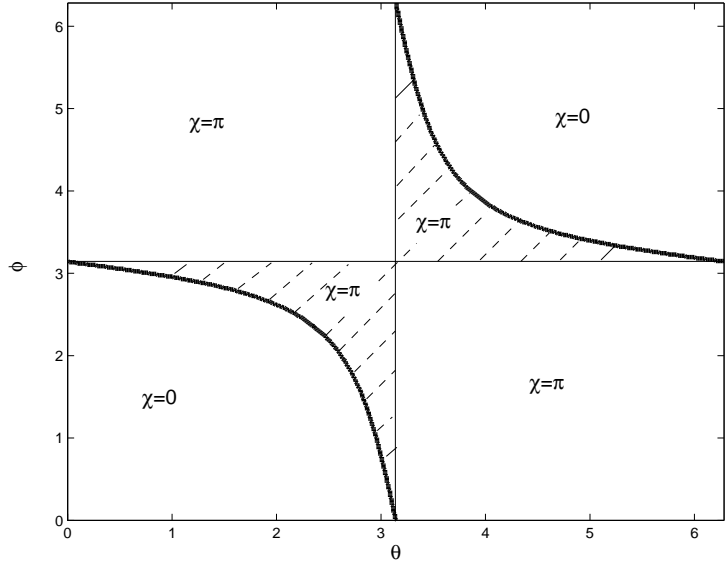}
 \end{center}
 \caption{{Phase diagram for the phase difference $\chi$ in the
$(\theta,\phi)$-plane. Solid line separates the regions with
$\chi=0$ and $\chi=\pi$. The dashed region is absent in the case
of local coupling.}}
 \label{chi}
\end{figure}

\begin{figure}[t]
 \begin{center}
 \includegraphics[width=0.8\textwidth]{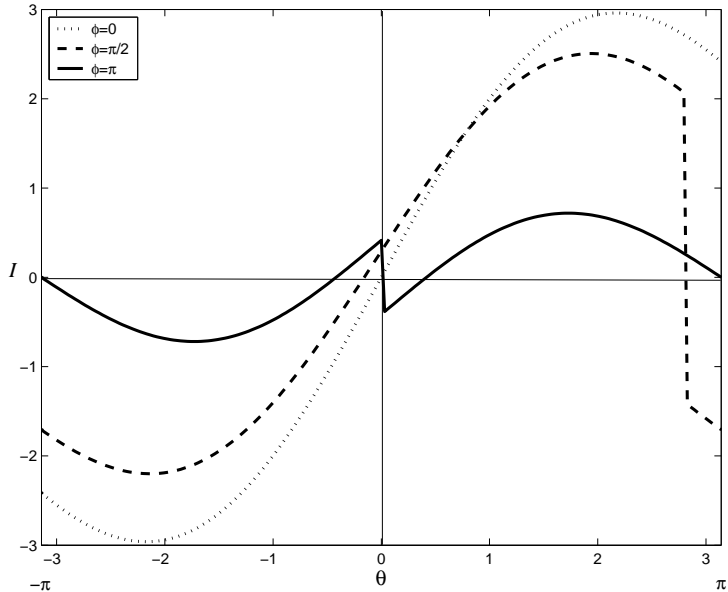}
 \end{center}
 \caption{{Current-phase relations $I(\theta)$ for different
values of $\phi$.}}
 \label{cur}
\end{figure}

The current-phase relations (11) and (12) with the condition (13)
are invariant under the transformation $\theta \rightarrow
\theta+2\pi n,$ and $\phi \rightarrow \phi+2\pi k$. The $2\pi$
periodicity of observable quantities is sustained by the ``hidden"
variable phase $\chi$. In Fig.3 the phase diagram for $\chi$ in
the $(\theta,\phi)$ plane is presented. The solid line separates
the regions with $\chi =0$ and $\chi =\pi$. When the state of the
system $(\theta,\phi)$ crosses this line, a jump in $\chi$ occurs.
Corresponding jumps take place in current-phase relations (11)
and (12). The current $I(\theta)$ (11) is shown in Fig.4 for
several values of the phase $\phi$. Note that the function
$I(\theta)$ has jumps, which for $\phi\neq 0$, are located not at
$\theta=\pm \pi$, as they would be in conventional 4-terminal
junctions. The jump in $\chi$ means the slippage of the phase
$\theta$ (or $\phi$). In the case of two-terminal or conventional
4-terminal junction the phase-slip events occur at phase
difference equal to $\pi(2n+1), n=0,\pm1,\pm2,..$. In
one-dimensional structures slippage of the phase occurs at
phase-slip centers (PSC), i.e. points where the order parameter
equals to zero. In our case of 2D mesoscopic 4-terminal weak
link, the analog of the PSC are phase-slip lines in normal metal
region. They appear when the state of system ($\theta,\phi$)
belongs to the dashed region in Fig.3. This region, which is
absent in the local coupling case (it actually coincides with
lines $\theta=\pi,\phi=\pi$), we call ``frustrated" region for
phases $\theta$ and $\phi$. For states inside this region, the
distribution of the supercurrent in the weak link contains 2D
vortex states (see below).

Nonlocal weak coupling  leads to the phase dragging effect
\cite{ommalek}. One notices that if $p\ne q$ then putting
$\theta=0$ in (11) results in a nonzero value of the current $I$
\begin{equation}
I = (p-q) \sin {\phi \over 2}\ {\rm sign} \left( \cos {\phi \over
2} \right)
\end{equation}
This current is absent in conventional 4-terminal junctions or
mesoscopic four terminal junctions with crossed implementation at
which $p=q$ (i.e. $det(\gamma_{\rm coupl})=0$).
\begin{figure}[t]
 \begin{center}
 \includegraphics[width=0.8\textwidth]{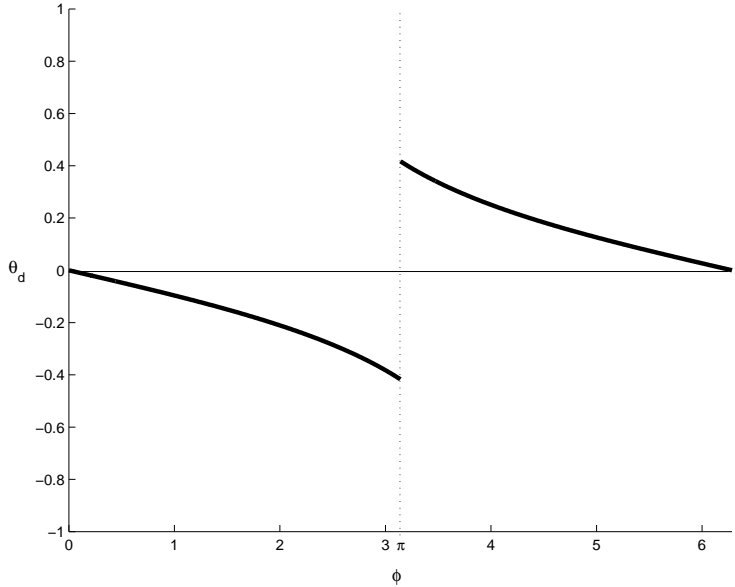}
 \end{center}
 \caption{{The dragged phase $\theta_d$ between terminals
S$_1$-S$_2$, at zero transport current $I$, as a function of the
phase difference $\phi$ between the other pair of terminals
S$_3$-S$_4$.}}
 \label{theta0}
\end{figure}

Similarly if we set $I=0$ in (11), we find a nonzero solution for
$\theta$, which again vanishes when $p=q$. This solution ($\equiv
\theta_d$) is a function of $\phi$ and is plotted versus $\phi$
in Fig.5. The influence of the phase of one side of the mesoscopic
4-terminal junction on the phase of the other side is what we call
phase dragging effect. This effect is one of important
characteristics of the  with parallel implementation.

In general current-phase relations are asymmetric, $I(-\theta) \ne
- I(\theta)$, unlike in conventional cases. In another words, the
presence of phase difference $\phi$ on the terminals S$_4$-S$_3$
breaks the time reversal symmetry for Josephson junction
S$_1$-S$_2$. It also follows from expression (11) that $I(\theta)$
is not only a function of $|\phi|$, as in conventional junctions,
but also depends on the sign of $\phi$. The phase dragging has
the analogy in the normal metal mesoscopic multiterminals,
described by formula (5); the normal current flowing through one
pair of terminals induces a voltage difference between the other
ones \cite{buttiker}.

\section{ Current distribution and local density of states
inside the mesoscopic weak link}

The coupling through the normal layer determines the behaviour of
the Josephson weak links S$_1$-S$_2$ and S$_4$-S$_3$. On the other
hand, the properties of the normal layer itself depend on the
phase differences $\theta$ and $\phi$ on the junctions. The phases
$\theta$ and $\phi$ can be controlled by external magnetic fluxes
through the rings (Fig.2a). In this section we present the
results of numerical calculations for current density
distribution ${\bf j} (\vec{\rho})$ and density of local Andreev
levels $N(\epsilon)$ inside the mesoscopic 4-terminal weak link.
The expressions for ${\bf j} (\vec \rho)$ and $N(\epsilon)$ as
functionals of $\{ \varphi_1,\varphi_2,\varphi_3,\varphi_4 \}$
were obtained in Ref.\cite{malekom} by solving Eilenberger
equations \cite{Eil}.
\begin{figure}[!t]
 \begin{center}
\includegraphics[totalheight=2 in]{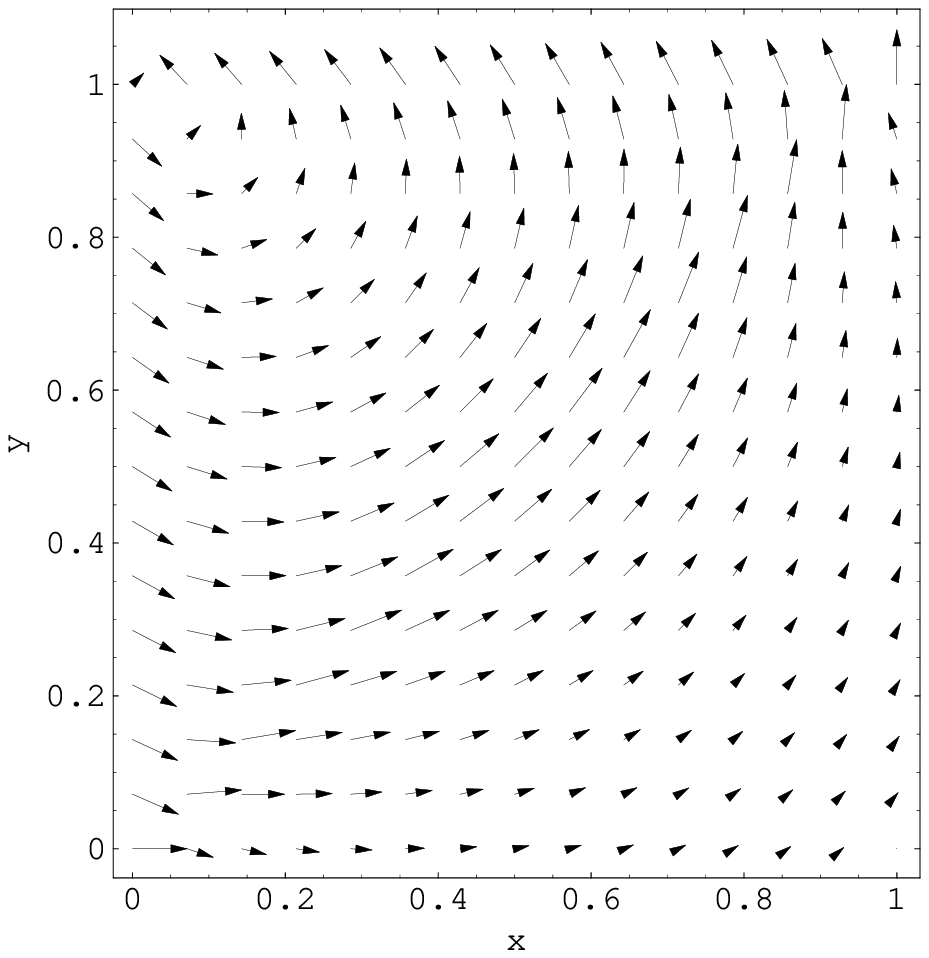}
\includegraphics[totalheight=2 in]{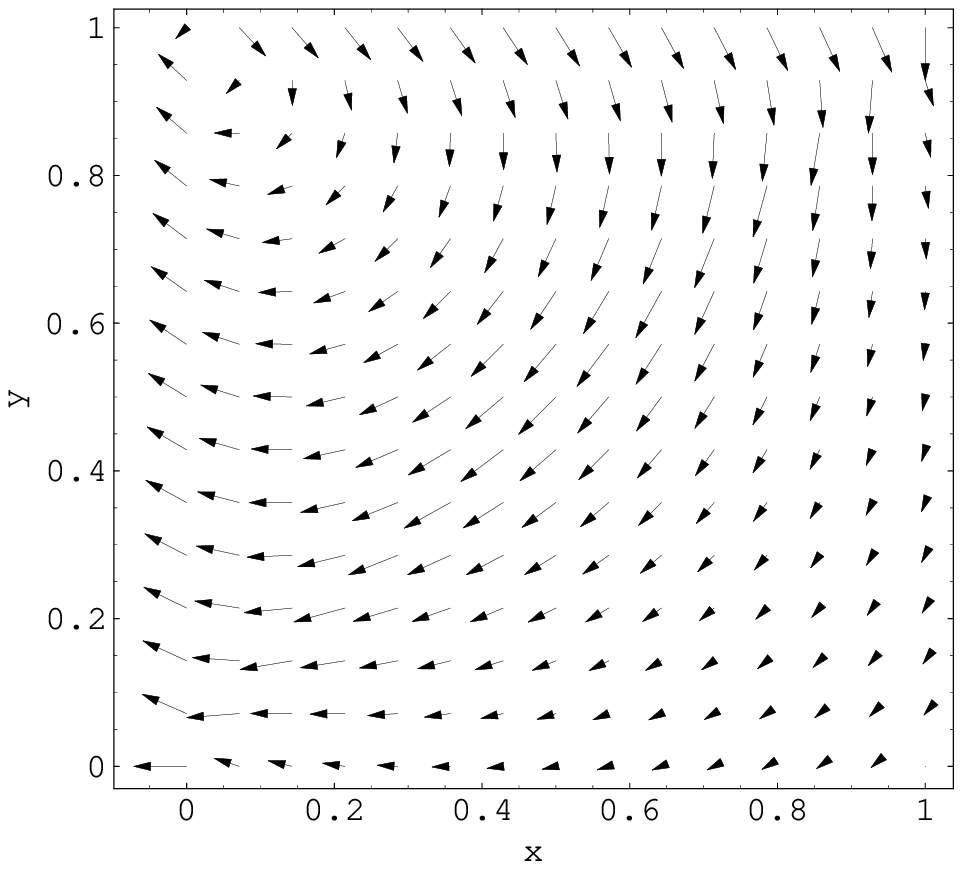}
\end{center}
\caption{{Distribution of the current density inside the normal
layer for phase $\phi=\pi$ and two values of the phase $\theta$
at which the current $I=0$ (Fig.4). (a) $\theta=-0.42$, (b)
$\theta=0.42$.}} \label{figure 6}
\end{figure}

Fig.6 illustrates the effect of phase dragging. Two sets of
phases ($\theta=-0.42$,\ $\phi=\pi$) and ($\theta=0.42$,\
$\phi=\pi$) correspond to zero value of the current $I$ (11) (see
Fig.4) and opposite directions of the current $J$ (12). In the
absence of the current from terminal S$_1$ to terminal S$_2$, the
phase difference on the junction S$_1$-S$_2$ exists.

When the phases $\theta$ and $\phi$ lie in the ``frustrated"
region of the diagram Fig.3 (dashed area), the current
distribution ${\bf j} (\vec{\rho})$ contains 2D vortex states.
They are shown in Fig.7 for states
$(\theta=\pi-0.2,\phi=\pi-0.2)$ and
$(\theta=\pi+0.2,\phi=\pi+0.2)$. In both cases, the order
parameter $\Psi(\vec \rho)$ vanishes along the diagonal $x=y$ and
its phase drops by $\pi$ when crossing this 2D phase-slip line.
\begin{figure}[!t]
 \begin{center}
\includegraphics[width=1\textwidth]{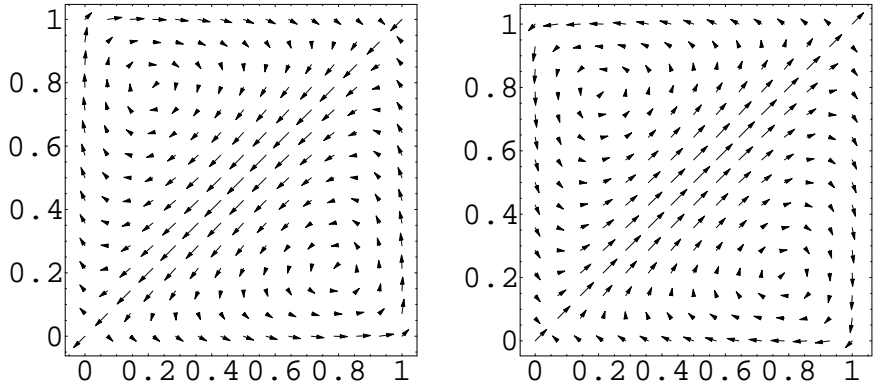}
\end{center}
\caption{{The vortex-like distributions of the current inside the
weak link when $\theta$ and $\phi$ are inside the frustrated
region: (a) $\theta=\phi=\pi-0.2$, (b) $\theta=\phi=\pi+0.2$ .}}
\label{Figure 7}
\end{figure}
\begin{figure}[!t]
 \begin{center}
\includegraphics[width=1\textwidth]{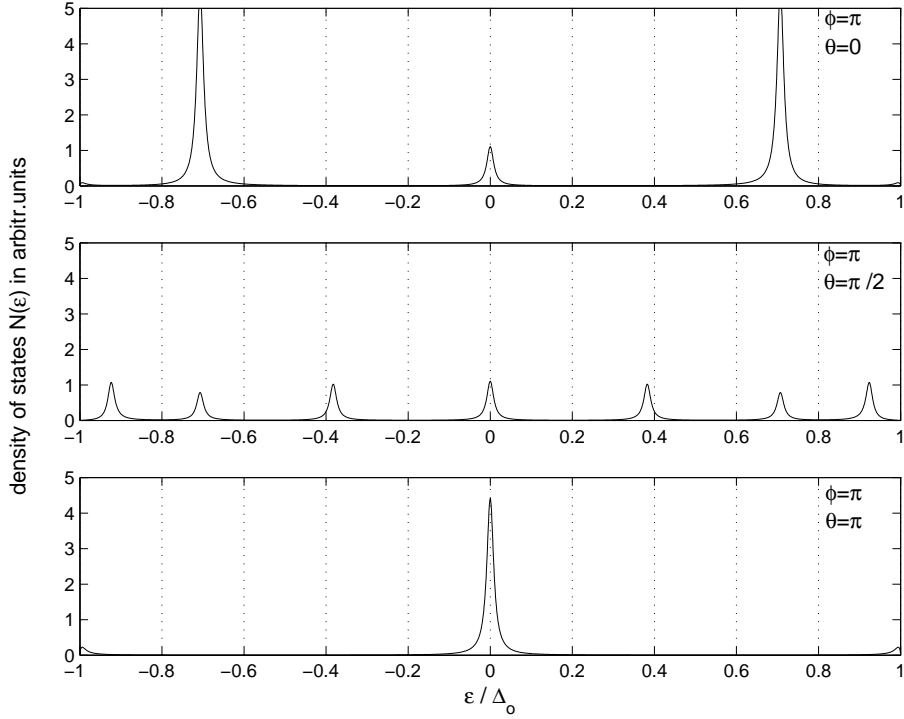}
\end{center}
\caption{{Density of states, $N(\epsilon)$, averaged over the
normal region for different values of $\theta$ and $\phi$.}}
\label{Figure 8}
\end{figure}

The Andreev scattering processes on the S$_i$-N interfaces lead
to the appearance of the energy levels with energies $\epsilon$
inside  the gap $\Delta_0$, $|\epsilon|< \Delta_0$, in the normal
metal. The local density of electron states in the normal layer is
given by the formula
\begin{equation}
{\cal N}(\epsilon,{\vec \rho})= N(0) <Re\ g(\omega=-i\epsilon,{\vec \rho},%
{\bf v}_F)>_{{\bf v}_F}.
\end{equation}
( $g(\omega,{\vec \rho},{\bf v}_F))$ is Eilenberger Green's
function). We have studied the dependence of the density of
states, averaged over area of the N-layer, $N(\epsilon)$, on the
phases $\theta$ and $\phi$. This tunneling density of states can
be measured by scanning tunneling microscope. It contains the
spikes with intensity and position on the energy axes controlled
by the phases $\theta$ and $\phi$. The results are shown in Fig.8
( the $\delta$-function singularities in $N(\epsilon)$ are smeared
by introducing a small damping $\Gamma=0.01 \Delta_0$).

\section{Mesoscopic Four-Terminal SQUID}

In this section we consider the four-terminal SQUID configuration
(Fig.2b). Conventional 4-terminal SQUID has been studied in
detail in Ref. \cite{oov}, wherein the steady states domain and
dynamical properties of the system were calculated. Here we are
interested in the specific features of the mesoscopic case
reflected in the current-phase relations (11,12). As we have seen
in the previous section, the nonlocal coupling ($ p\neq q $)
leads to the phase dragging effect. This dragged phase can induce
a transferred magnetic flux in the ring which depends on the
transport current. On the other side, the magnetic flux state in
the ring influences the behaviour of the Josephson junction in
the current circuit.

When the terminals 3 and 4 are short circuited by a
superconducting ring with self-inductance $L$, the phase $\phi$
is related to the observable quantity, magnetic flux threading the
ring $\Phi$, $\phi=\frac{2e}{\hbar}\Phi $. The current $J$
circulating in the ring is given by $J=(\Phi^e-\Phi)/L$, where
$\Phi^e$ is the external magnetic flux threading the ring. From
(11) and (12) we have
\begin{equation}
I=\sin{\theta}+
\left[(p+q)\sin{\frac{\theta}{2}}\cos{\frac{\Phi}{2}}+
(p-q)\cos{\frac{\theta}{2}}\sin{\frac{\Phi}{2}} \right]\cos{\chi},
\end{equation}
\begin{equation}
\frac {\Phi^e-\Phi} {{\cal L}}=\sin{\Phi}+\frac{1}{\kappa}
\left[(p+q)\sin{\frac{\Phi}{2}}\cos{\frac{\theta}{2}}+
(p-q)\cos{\frac{\Phi}{2}}\sin{\frac{\theta}{2}} \right]\cos{\chi},
\end{equation}
where fluxes $\Phi,\Phi^e$ are measured in units $\hbar/2e $,
${\cal L}= (2e/\hbar)LI_0\kappa$ is the dimensionless
self-inductance. The parameter $\kappa=\gamma_{34}/\gamma_{12}$
is the ratio of the critical currents of the sub-junctions $3-4$
and $1-2$. The limiting cases of $\kappa \rightarrow \infty$ and
$\kappa \rightarrow 0$ correspond to the autonomous SQUID and the
current biased Josephson junction, respectively.

The transport current $I$ and the external flux $\Phi^e$ are the
external controlling parameters. The corresponding Gibbs
potential for the 4-terminal SQUID takes the form
\begin{eqnarray}
  && U(\Phi,\theta;I,\Phi^e) = \frac {\kappa (\Phi-\Phi^e)^2} {2\cal
  L} - I \theta -\cos(\theta)- \kappa \cos(\Phi) \nonumber \\
  &&- 2 \left((p+q) \cos{\theta\over 2}\cos {\Phi \over 2}-(p-q)
  \sin {\theta\over 2} \sin {\Phi\over 2} \right) \cos(\chi)
\end{eqnarray}
The last three terms in Eq.(18) are the Josephson coupling energy
(4) in terms of variables $\theta$ ,$\Phi$ and $\chi$. The
minimization of $U$ with respect to $\chi$ gives the expression
(13) for $\cos(\chi)$, with $\phi$ replaced by $\Phi$. At given
values of the control parameters $I$ and $\Phi^e$, the relations
(16) and (17) (together with Eq.(13)) determine the set of
possible states of the system $\{\theta,\Phi\}$, among which we
should choose those that correspond to the local minima of the
potential $U$, Eq.(18).
\begin{figure}[!t]
 \begin{center}
\includegraphics[width=0.8\textwidth]{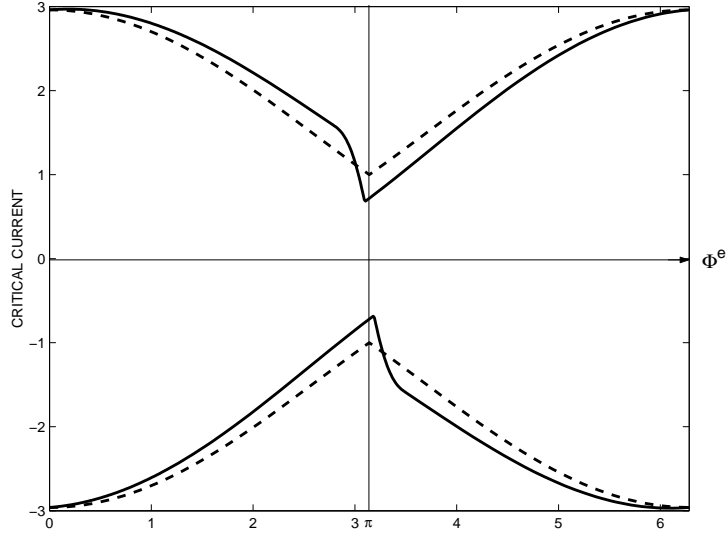}
\end{center}
\caption{{The steady state domain for mesoscopic four-terminal
SQUID in plane $(I,\Phi_e)$ of the control parameters (solid
line). Dashed line corresponds to the conventional four-terminal
SQUIDs.}} \label{Figure 9}
\end{figure}
\begin{figure}[!t]
 \begin{center}
\includegraphics[width=0.7\textwidth]{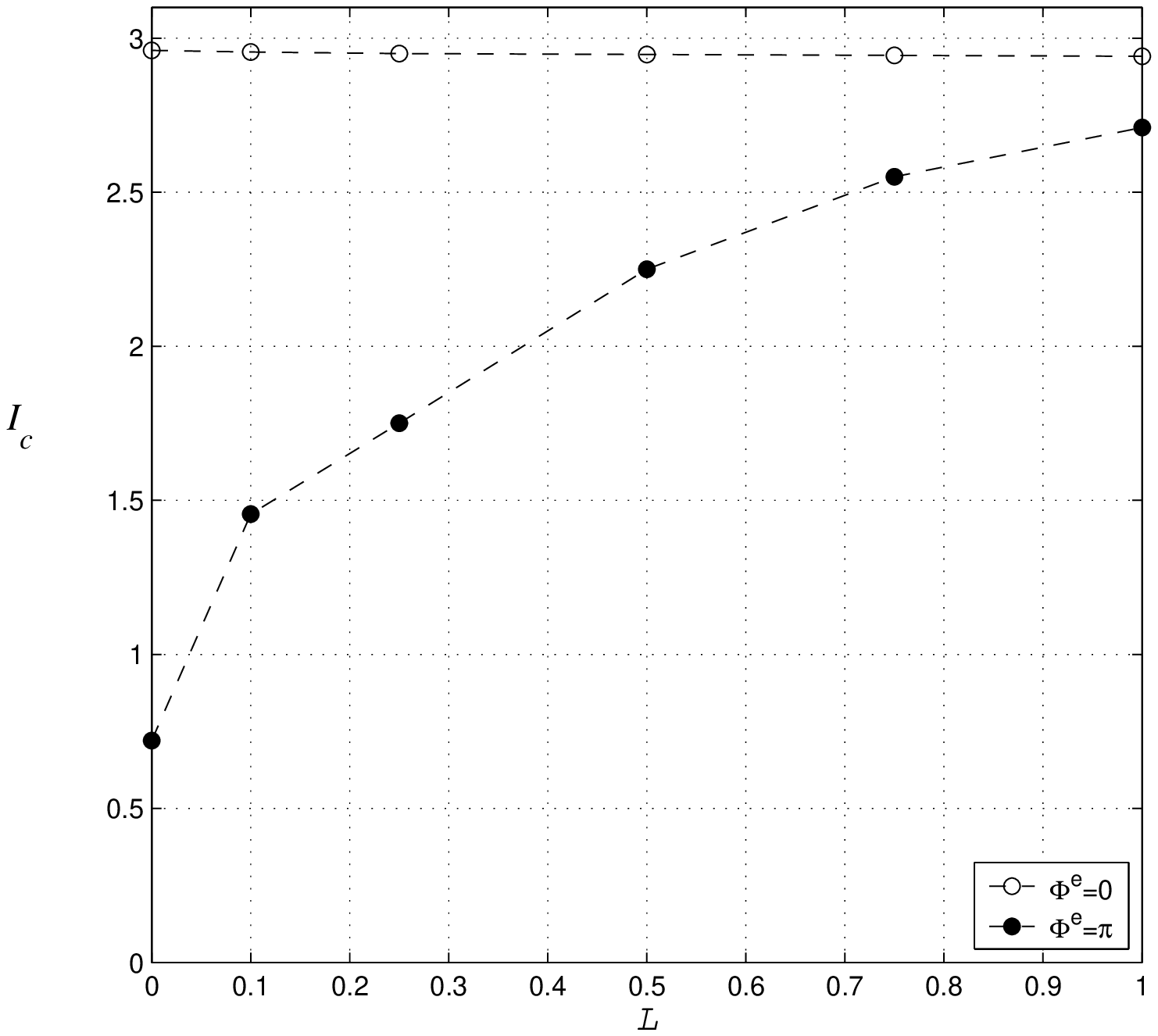}
\end{center}
\caption{{The critical current, $I_c$, between the superconductors
S$_1$ and S$_2$, as a function of ${\cal L}$ for $\Phi_e=0$ and
$\pi$.}} \label{Figure 10}
\end{figure}
Let us consider the effect of the magnetic flux state of the ring
on the behaviour of the current driven junction. The critical
current of the junction, $I_c$, depends on the applied magnetic
flux $\Phi^e$. In the simplest case of small self-inductance
${\cal L} \ll1$, we can neglect the difference between $\Phi$ and
$\Phi^e$ in expression (16).

The maximal value of the supercurrent $I$ (16) (with $\Phi$
replaced by $\Phi^e$) as a function of $\Phi^e$,
$I_{max}(\Phi^e)$, is shown in Fig.9. This curve determines the
boundary of the steady states domain in the $(I,\Phi^e)$ plane.
The function $I_{max}(\Phi^e)$ is $2\pi$ periodic, but due to the
terms proportional to $p-q$ in Eq.(16), it is not invariant under
the transformation $\Phi^e \rightarrow -\Phi^e$. The symmetry is
restored if we simultaneously change $\Phi^e$ on $-\Phi^e$ and
$I$ on $-I$. Note, that in conventional case ($p=q$) the boundary
of the steady state domain $I_{max}(\Phi^e)$ is symmetric with
respect to the axes $(I,\Phi^e)$ (dashed line in Fig.9). Thus,
the critical current $I_c$ in the transport current circuit, for
a given direction of the current, depends on the sign of the
magnetic flux in the ring. For finite values of self-inductance
${\cal L}$ equations (16) and (17) must be treated
self-consistently. The critical current $I_c$ as function of
${\cal L}$ is shown in Fig.10 for two values of external flux,
$\Phi^e = 0$ and $\Phi^e = \pi$.

Outside the steady state domain, the stationary solutions for
$(\theta,\Phi)$ are absent and system goes to the nonstationary
resistive regime. The simple generalization of Eqs. (16,17) in the
frame of the heavily damped resistively shunted junction (RSJ)
model \cite{bar} leads to equations (see \cite{oo}):
\begin{equation}
\frac{d\theta}{dt}=I-\sin{\theta}-
\left[(p+q)\sin{\frac{\theta}{2}}\cos{\frac{\Phi}{2}}+
(p-q)\cos{\frac{\theta}{2}}\sin{\frac{\Phi}{2}} \right]\cos{\chi},
\end{equation}
\begin{equation}
\frac{d\Phi}{dt}= \frac {\Phi^e-\Phi} {{\cal
L}}-\sin{\Phi}-\frac{1}{\kappa}
\left[(p+q)\sin{\frac{\Phi}{2}}\cos{\frac{\theta}{2}}+
(p-q)\cos{\frac{\Phi}{2}}\sin{\frac{\theta}{2}} \right]\cos{\chi},
\end{equation}
\begin{equation}
\frac{d\chi}{dt}=-\sin(\chi) \left[(p+q)\cos{\frac{\Phi}{2}}
\cos{\frac{\theta}{2}}-(p-q)\sin{\frac{\Phi}{2}}
\sin{\frac{\theta}{2}} \right].
\end{equation}
They can also be presented in a form
\begin{equation}
\dot \theta =-\frac{\partial U}{\partial \theta }\ , \quad \dot
\Phi =- \frac{\partial U}{\partial \Phi }\ , \quad \dot \chi
=-\frac {1}{2}\frac{\partial U}{\partial \chi },
\end{equation}
where potential $U$ is defined in Eq.(18). The voltages between
different terminals are related to the time derivatives of the
phase differences
\begin{equation}
V_{21}=\dot \theta , \quad V_{34}=\dot \Phi , \quad \frac
12(V_{13}+V_{24})=\dot \chi .
\end{equation}
The time and the voltage are measured in the units of $e/I_0$ and
$\hbar I_0/2e^2$ respectively. Note that, inspite of the
equilibrium state, the dynamical variable $\chi$ relates to an
observable quantity. It's time derivative determines the voltage
between the ring and the transport circuit. The features of the
dynamical behaviour of the mesoscopic 4-terminal SQUID are again
affected by the terms proportional to $(p-q)$, i.e. by nonlocal
coupling. The current-voltage characteristics in the transport
channel, $V(I)$, (the time averaged voltage $V_{21}$ (23)), can be
obtained by the numerical solution of the coupled system of
nonlinear differential equations (20-22). As well as critical
current $I_c$, the voltage $V(I)$ in applied magnetic flux
$\Phi^e$ depends on the sign of the $\Phi^e$, i.e on the
direction of the external magnetic field. Full dynamical
description of the mesoscopic four-terminal SQUID requires more
rigorous approach than RSJ model, and will be the the subject of
separate investigation.
\begin{figure}[!t]
 \begin{center}
\includegraphics[width=0.7\textwidth]{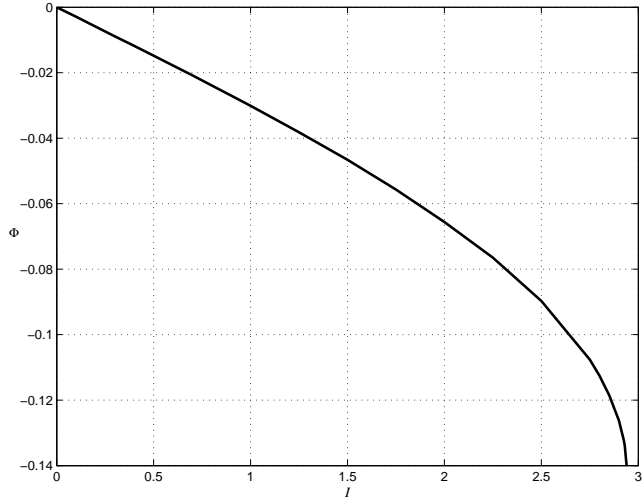}
\end{center}
\caption{{The flux induced inside the ring as a function of the
transport current $I$. ${\cal L}=1$,$\Phi_e=0$}} \label{Figure 11}
\end{figure}
In accordance with stationary (16,17) or dynamical equations
(19-21) for $\theta$ and $\Phi$ the opposite effect for influence
of transport current circuit on the flux states in the ring takes
place. In particular, the current $I$ produces the flux $\Phi$ in
the ring even in stationary case and in the absence of external
flux $\Phi^e$. This effect is proportional to $(p-q)$ and is
absent in the conventional case. In Fig.11 we plot the dependence
of the induced in the ring magnetic flux $\Phi$ on the transport
current $I$ in the case $\Phi^e = 0$.

Special interest presents the existence of the bistable states in
the system described by the potential (18). We emphasize, that in
contrast to the usual SQUID, bistable states occur for any
inductance ${\cal L}$, even for ${\cal L}< 1$ \cite{oov}. We will
analyze the dependence of these states on the control parameters
$I$ and $\Phi^e$ in article II, when designing of the
four-terminal qubit will be studied.

\section{Conclusions}

We have demonstrated that in ballistic four-terminal Josephson
junctions coherent anisotropic superposition of macroscopic wave
functions of the superconductors in the normal region produces
formation of phase slip lines (2D analogs to phase slip centres)
and time-reversal symmetry breaking 2D vortex states in it, as
well as such effects as phase dragging and magnetic flux
transfer. We have calculated the phase-dependent tunneling
density of Andreev states in this region as well.

A degree to which the nonlocality of mesoscopic transport is
manifested, depends on the characteristics of the system and is
most pronounced in the ballistic case \cite{KOS}. Ballistic
four-terminal junctions considered here demonstrate several
specific effects absent in the diffusive limit \cite{VO1,oov,AL}:
the phase dragging, time-reversal symmetry breaking
($I(\theta)\neq I(-\theta)$, Eq.(14)), and the vortex formation.
The latter can mimic the behaviour of SNS junctions with
unconventional superconductors \cite{Huck}. It has indeed the
same origin in direction-dependent phase of the superconducting
order parameter induced in the normal part of the system, though
not due to the intrinsic phase difference between different
directions in a superconductor. This actually allows us more
freedom in controlling the behaviour of the junction, which will
be exploited in the qubit design based on such a junction in the
following paper. The time-reversal symmetry breaking can be also
used for direction-sensitive detection of weak magnetic fluxes.

It will be instructive to investigate the role played by finite
elastic scattering in the system and look for the analogs of zero
bound states, found at surfaces/interfaces of unconventional
superconductors (for a review see \cite{KT}). This, as well as
vortex dynamics in the system, will be the subject of our further
research.

\textbf{Acknowledgements}

We thanks R.de.Bruyn Ouboter for his stimulating interest in this
work. One of the authors, A.N.O., would like to acknowledge the
D-Wave Systems Inc. (Vancouver) for hospitality and support for
this research.

\section{Appendix. Junction with arbitrary $\gamma$'s}

The Josephson energy of the mesoscopic Four Terminal Junction,
normalized to $(\hbar/2e) (\pi {\Delta_0(T)}^2 / 4 e T_c)$, is
expressed by
\begin{equation}
E_J= -\gamma_{12} \cos \theta - \gamma_{34} \cos \phi + E_{\rm
coupl}
\end{equation}
with the coupling energy $E_{\rm coupl}$ given by
\begin{eqnarray}
E_{\rm coupl} &=& - \gamma_{13} \cos \left({-\theta-\phi \over 2}
+ \chi \right) - \gamma_{14} \cos \left({-\theta+\phi
\over 2} + \chi \right) \nonumber \\
&& - \gamma_{23} \cos \left({\theta-\phi \over 2} + \chi \right)
- \gamma_{24} \cos \left({\theta+\phi \over 2} + \chi \right)
\nonumber \\
&=& - (A \cos \chi + B \sin \chi)
\end{eqnarray}
where
\begin{eqnarray}
&& A= (\gamma_{13} + \gamma_{24}) \cos \left({\theta+\phi \over
2}\right) + (\gamma_{14} + \gamma_{23}) \cos \left({\theta-\phi
\over 2}\right) \nonumber \\
&& B= (\gamma_{13} - \gamma_{24}) \sin \left({\theta+\phi \over
2}\right) + (\gamma_{14} - \gamma_{23}) \sin \left({\theta-\phi
\over 2}\right)
\end{eqnarray}
Minimizing with respect to $\chi$, we find the minimum to be
\begin{equation}
E_{\rm coupl} = - \sqrt{A^2+B^2}, \qquad \chi = \cos^{-1} \left(
{A \over \sqrt{A^2+B^2}} \right) \label{Echi}
\end{equation}
After some manipulations we find
\begin{eqnarray}
E_{\rm coupl} = &-&[ \ \gamma_{13}^2 + \gamma_{23}^2 +
\gamma_{14}^2 + \gamma_{24}^2 \label{Ecoupl} \\ && +
2(\gamma_{13}\gamma_{14} + \gamma_{23}\gamma_{24}) \cos\phi +
2(\gamma_{13}\gamma_{23}  +
\gamma_{14}\gamma_{24}) \cos\theta \nonumber \\
&& + 2(\gamma_{13}\gamma_{24} + \gamma_{14}\gamma_{23})\cos\theta
\cos\phi - 2(\gamma_{13}\gamma_{24} -
\gamma_{14}\gamma_{23})\sin\theta \sin\phi \ ]^{1/2}\nonumber
\end{eqnarray}
The last term in the bracket in (\ref{Ecoupl}) vanishes when \
$det(\gamma_{coupl})=0$. In that case the current $I(\theta,\phi)$
will be zero at $\theta=0$. On the other hand, if \
$det(\gamma_{coupl}) \ne 0$ then $I(\theta,\phi)\ne 0$ when
$\theta=0$. This is a signature of the phase dragging effect.

In a four terminal junction with micro-bridges near $T_c$ one has
$\gamma_{ij} \sim 1/R_i R_j$. In that case the last term in
(\ref{Ecoupl}) will vanish and $E_{\rm coupl}$ factorizes
\begin{equation}
E_{\rm coupl} \sim - \left[ \left({1\over R_1} - {1\over R_2}
\right)^2 + {4\cos^2 (\theta/2) \over R_1 R_2} \right]^{1\over 2}
\left[ \left({1\over R_3} - {1\over R_4} \right)^2 + {4\cos^2
(\phi / 2)\over R_3 R_4} \right]^{1\over 2}
\end{equation}
In particular when $R_1=R_2$ and $R_3=R_4$ we find
\begin{equation}
E_{\rm coupl} \sim - {4\over R_1R_3}\left| \cos {\theta \over 2}
\right| \left| \cos {\phi \over 2} \right|
\end{equation}
which is what one obtains from Ginzburg-Landau calculation.

In a mesoscopic four terminal junction with parallel
implementation on the other hand, we have
$\gamma_{13}=\gamma_{24}$ and $\gamma_{14}=\gamma_{23}$. This
leads to $B=0$ and therefore
\begin{equation}
E_{\rm coupl} = - |A|, \qquad \cos \chi= {\rm sign} (A)
\end{equation}
which gives $\chi= 0$ or $\pi$. Notice that in the general case of
(\ref{Echi}), $\chi$ can take other values than $0$ and $\pi$.

\end{document}